\documentclass[conference]{IEEEtran}
\IEEEoverridecommandlockouts % for the thanks comment

\usepackage{graphicx}      % include this line if your document contains figures

\usepackage{commath}
\usepackage{amsfonts}
\usepackage{mathtools}
\usepackage{steinmetz}		% for phasor
\usepackage{supertabular}	% for long tabular
\usepackage{tabularx}		% for long text in tabular
\usepackage{algorithm}		% for algori1thms
\usepackage{algorithmic}		% for algorithms
\usepackage{amsmath,amssymb}
\usepackage{amsthm}
\newtheorem{rem}{Remark}
\newtheorem{defi}{Definition}
\newtheorem{prop}{Proposition}

\theoremstyle{proof}

\usepackage{subfig}
\usepackage{enumitem} 		% for itemize margins
\usepackage{pgfplots}
\pgfplotsset{width=8cm,height=4cm,compat=1.9}

\usepackage{soul}

\renewcommand{\b}{\textcolor{black}}

\usepackage{cite}
\ifCLASSINFOpdf
\else
\fi
\usepackage{array}
\begin{document}
\title{Efficient Convex Optimization for Optimal PMU Placement in Large Distribution Grids}

%\author{\IEEEauthorblockN{
%Miguel Picallo\IEEEauthorrefmark{1}, 
%Adolfo Anta\IEEEauthorrefmark{2}}
%\IEEEauthorblockA{\IEEEauthorrefmark{2}General Electric Global Research \\
%Munich, Germany \\
%\{anta,panosyan\}@ge.com}
%\and
%\IEEEauthorblockN{Bart De Schutter\IEEEauthorrefmark{1}}
%\IEEEauthorblockA{\IEEEauthorrefmark{1}Delft Center for Systems and Control \\
%Delft University of Technology \\
%Delft, The Netherlands \\
%\{M.picallocruz,b.deschutter@tudelft.nl\}}
%\thanks{This project has received funding from the European Union's Horizon 2020 research and innovation programme under the Marie Skł{\l}odowska-Curie grant agreement No 675318 (INCITE).}
%}

\author{\IEEEauthorblockN{Miguel Picallo\IEEEauthorrefmark{1},
Adolfo Anta\IEEEauthorrefmark{2},
Bart De Schutter\IEEEauthorrefmark{1}}
 \IEEEauthorblockA{\IEEEauthorrefmark{1}Delft Center for Systems and Control, Delft University of Technology \\
Delft, The Netherlands \\
\{m.picallocruz,b.deschutter\}@tudelft.nl}
\IEEEauthorblockA{\IEEEauthorrefmark{2}Austrian Institute of Technology \\
Vienna, Austria \\
Adolfo.Anta@ait.ac.at}
\thanks{This project has received funding from the European Union's Horizon 2020 research and innovation programme under the Marie Skł{\l}odowska-Curie grant agreement No 675318 (INCITE).}%
}

\maketitle

\begin{abstract}
The small amount of measurements in distribution grids makes their monitoring difficult. Topological observability may not be possible, and thus, pseudo-measurements are needed to perform state estimation, which is required to control elements such as distributed generation or transformers at distribution grids. Therefore, we consider the problem of optimal sensor placement to improve the state estimation accuracy in large-scale, 3-phase coupled, unbalanced distribution grids. This is an NP-hard optimization problem whose optimal solution is unpractical to obtain for large networks. For that reason, we develop a computationally efficient convex optimization algorithm to compute a lower bound on the possible value of the optimal solution, and thus check the gap between the bound and heuristic solutions. We test the method on a large test feeder, the standard IEEE 8500-node, to show the effectiveness of the approach.
\end{abstract}



% \fi
\IEEEpeerreviewmaketitle

\section{Introduction}
\vspace{-0.01cm}
Accurate monitoring of voltages, currents, and loads is essential to manage an electrical power network. In this context, State Estimation (SE) consists in estimating the network state, represented typically by the bus voltage phasors. Normally, SE computes the state that best concurs with the available measurements by solving a weighted least-squares problem using an iterative approach like Newton-Raphson \cite{abur2004power, monticelli2000electric}. 
%In opposite to transmission networks, which have a meshed structure and distributed generation injecting power all over the network; distribution grids typically have a radial structure and used to have a single source node injecting power. However, distribution grids are changing due to the increasing deployment of distributed generation like PV panels, batteries, etc.; and therefore, SE is becoming necessary \cite{ipakchi2009grid}.

Network monitoring can be improved with the introduction of Phasor Measurements Units (PMUs). Some previous work proposes to achieve topological observability \cite{baldwin1993power} by using integer linear programming \cite{gou2008generalized}. However, although cheap PMUs are becoming available \cite{pinte2015mpmu}, their operational and network communication costs\cite{beg2016pmucom} may still prevent installing enough units to guarantee topological observability and thus also numerical observability, which is necessary for SE \cite{baldwin1993power}. Therefore, PMUs need to be combined with Supervisory Control And Data Adquisition (SCADA) measurements to solve the SE problem \cite{zhou2006alternative}. Recent work \cite{kekatos2012optimal} presents the problem of optimal PMU placement as the minimization of the error covariance matrix in SE through some metric from optimal design of experiments \cite{pukelsheim2006optimal}. This transforms the optimal PMU placement problem into a combinatorial optimization problem with binary variables and a nonlinear objective function. As a consequence, as the network becomes larger, the optimal solution cannot be computed within a reasonable time span, due to the number of possible combinations of measurements.

The problem is even more complex in distribution grids, where even when SCADA measurements are used, the number of measurements is not enough to achieve observability. As a consequence, SE for distribution grids needs to rely on so-called pseudo-measurements, like load forecasts, which may have a high uncertainty associated to them (approx. $50$\% relative error \cite{schenato2014bayesian}). This motivates the use of PMUs in distribution grids \cite{meier2014pmudg} to increase the SE accuracy, by for example using greedy and random combinatorial approaches of sensors \cite{singh2009measurement, singh2011meter} or evolutionary algorithms \cite{liu2012tradeoff, Prasad2018tradeoff}. However, these methods do not provide optimality guarantees, nor do they inform how good their suboptimal solutions are with respect to the optimal one. A convexity-based lower bound of the value of the optimal solution can provide this information \cite{kekatos2012optimal}, but this is difficult to compute for large grids.

Our main contribution is to propose a computationally efficient and scalable approach to derive this convexity-based lower bound for the optimal solution of the PMU placement problem under a budget constraint in large distribution grids. In particular, we extend the projection algorithm in \cite{insense2018} to the budget-constrained case by using the Karush-Kuhn-Tucker conditions. Then, we derive subgradient expressions for the A,D,E,M-optimal metrics \cite{pukelsheim2006optimal} and combine them with the projection algorithm to create a projected subgradient descent method to solve the convex relaxed optimal PMU placement problem in large distribution grids. 

The rest of the paper is structured as follows: Section \ref{sec:grid} presents some background about power networks. Section \ref{sec:info} discusses the different types of measurements. Section \ref{sec:SEbasic} summarizes the newly proposed methodology for SE \cite{picallo2017twostepSE}. Section \ref{sec:optplac} presents the metrics considered and states the problem of optimal sensor placement problem under a budget constraint. Section \ref{sec:opt} shows the optimization algorithms used to solve the problem. Section \ref{sec:testcase} shows the results on a test case of a large distribution network. Finally, Section \ref{sec:conc} presents the conclusions.

%\section{Nomenclature}
%\begin{tabular}{ll}
%$V$ & Vector of bus voltages\\
%$I$ & Vector of bus currents\\
%$S$ & Vector of bus loads\\ 
%$Y$ & Admittance matrix \\
%$V_0$ & Vector of voltages at 0 loads \\
%$S_\text{psd}$ & Pseudo-measurement estimations of loads \\ 
%$\varepsilon$ & Indices of buses with no load connected \\
%$z_\text{meas}$ & Real-time phasor measurements \\ 
%$z_\text{measNL}$ & Real-time magnitude measurements \\ 
%\end{tabular}

\section{Distribution Grid Model}\label{sec:grid}
\vspace{-0.01cm}
% graph
A distribution grid consists of buses, where power is injected or consumed, and branches, each connecting two buses. This system can be modeled as a graph $\mathcal{G}=(\mathcal{V},\mathcal{E},\mathcal{W})$ with nodes $\mathcal{V}=\{1,...,N_\text{bus}\}$ representing the buses, edges $\mathcal{E}=\{(v_i,v_j)\mid v_i,v_j \in \mathcal{V}\}$ representing the branches, and edge weights $\mathcal{W}=\{w_{i,j}\mid (v_i,v_j) \in \mathcal{E}\}$ representing the admittance of the branches, which are determined by the length and type of the line cables.

%3-phase
In 3-phase networks buses may have up to 3 phases, so that the voltage at bus $i$ is $V_{\text{bus},i} \in \mathbb{C}^{n_{\phi,i}}$, where $n_{\phi,i}\leq 3$ (and the edge weights $w_{i,j}\in \mathbb{C}^{n_{\phi,i} \times n_{\phi,j}}$). The state of the network is then typically represented by the vector bus voltages $V_\text{bus}=[V_\text{src}^T, \; V^T]^T \in \mathbb{C}^{N+3}$, where $V_{\text{src}} \in \mathbb{C}^3$ denotes the known voltage of the 3 phases at the source bus, and $V \in \mathbb{C}^N$ the voltages in the non-source buses, where $N$ depends on the number of buses and phases per bus.

% power flow
Using the Laplacian matrix $Y \in \mathbb{C}^{(N+3) \times (N+3)}$ of the weighted graph $\mathcal{G}$, called admittance matrix \cite{abur2004power}, the power flow equations to compute the currents $I$ and the power loads $S$ are:
\begin{equation}\label{eq:PFeq}\arraycolsep=1pt
\begin{array}{c}
\left[\begin{array}{c} I_{\text{src}} \\ I \end{array}\right] = 
Y\left[\begin{array}{c} V_{\text{src}} \\ V \end{array}\right], \; S = \text{diag}(\bar{I})V
\end{array}
\end{equation}
where $\bar{(\cdot)}$ denotes the complex conjugate, $\text{diag}(\cdot)$ represents the diagonal operator, converting a vector into a diagonal matrix. 
%\y{Separating $Y$ in blocks according to the indices of the source bus $V_{\text{src}}$, see \eqref{eq:PFeq}, the voltage $V$ for the non-source buses can be rewritten as:
%\begin{equation}\label{eq:PFit}
%V =  Y_\text{d}^{-1}I-Y_\text{d}^{-1}Y_\text{c}V_{\text{src}}, \; V_0=V\mid_{I=0}=-Y_\text{d}^{-1}Y_\text{c}V_{\text{src}}
%\end{equation}
%with $V_0$ denoting the voltage without loads.}

\section{Measurements} \label{sec:info}
\vspace{-0.01cm}
As explained in \cite{picallo2017twostepSE}, several different sources of information can be available to solve the SE problem:
\begin{enumerate}[leftmargin=*]
\item \textit{Pseudo-measurements}, i.e., load estimations $S_\text{psd}$ based on predictions and/or known installed load capacity at every bus. Since these pseudo-mearurements are estimations, they are modeled as noisy measurements with a Gaussian distribution with a relative large standard deviation (a typical value can be $\sigma_\text{psd} \approx 50\%$ \cite{schenato2014bayesian}).

\item \textit{Virtual measurements}, i.e., buses with zero-injections, no loads connected. They can be modeled as physical constraints for the voltage states by defining the set of indices of zero-injection buses $\varepsilon=\{i,\cdots,j\}$:
\begin{equation}\label{eq:Scons}
(S)_{\varepsilon}=0, \; (I)_{\varepsilon}=0
\end{equation} 
where $(\cdot)_{\varepsilon}$ denotes the elements at indices in ${\varepsilon}$.

\item \textit{Real-time PMU measurements}, i.e., voltage and current GPS-synchronized measurements of magnitude and phase angle. According to the IEEE standard for PMU \cite{martin2008exploring}, we model the noises with a low standard deviation for the magnitude and the angle, $\sigma_\text{mag} \approx 1\%$ and $\sigma_\text{ang} \approx 0.01\text{ rad}$ respectively. They can be expressed using a linear approximation \cite{picallo2017twostepSE} with magnitude and angle noise due to the measurements and imperfect synchronization. For a number of $N_\text{meas}$ measurements $z_\text{meas} \in \mathbb{C}^{N_\text{meas}}$ we have:
\begin{equation}\label{eq:Lmeas}
\begin{array}{c}
z_\text{meas} \approx  C_\text{meas}V + \text{diag}(C_\text{meas}V)(\omega_{\text{mag}} + j\omega_{\text{ang}})
\end{array}
\end{equation}
where $\omega_\text{mag} \hspace{-0.1cm} \sim \hspace{-0.1cm} \mathcal{N}(0,\sigma_\text{mag}I_{\text{d},N_\text{meas}})$, $\omega_\text{ang} \hspace{-0.1cm} \sim \hspace{-0.1cm} \mathcal{N}(0,\sigma_\text{ang}I_{\text{d},N_\text{meas}})$, with $I_{\text{d},n}$ denoting the identity matrix of dimension $n$. $C_\text{meas}$ is the matrix mapping state voltages to measurements. Then, for measurement $j$ at phase $l$ of bus $i$ we have:
\begin{equation}\label{eq:LmeasMap}\arraycolsep=1pt\begin{array}{l}
(C_\text{meas}V)_j = (C_\text{meas})_{j,\bullet}V= \\[0.1cm]
\left\lbrace \begin{array}{ll}
V_{i_l} & \mbox{for a voltage measurement} \\[0.0cm]
(Y)_{i_l,\bullet}V & \mbox{for a current measurement} \\[0.0cm]
(Y)_{i_l,m_l}(V_{i_l}-V_{m_l}) & \mbox{for a branch-current $i \to m$} \\[-0.0cm]
&  \mbox{measurement}
\end{array} \right.
\end{array}
\end{equation}
where $(\cdot)_{j,\bullet}$ denotes row $j$. Since the measurement noises in \eqref{eq:Lmeas} are small according to the PMUs standard \cite{martin2008exploring}, their covariance matrices can be approximated using the measurements:
\b{
\begin{equation*}
\begin{array}{rl}
\Sigma_\text{meas} & = (\sigma_\text{mag}^2+\sigma_\text{ang}^2) \text{diag}(\:\abs{C_\text{meas}V}^2) \\
& \approx (\sigma_\text{mag}^2+\sigma_\text{ang}^2) \text{diag}(\:\abs{z_\text{meas}}^2)
\end{array}
\end{equation*}
}
\end{enumerate}

\section{State Estimation}\label{sec:SEbasic}
\vspace{-0.01cm}
Typically, SE consists in finding the voltages that best match the measurements by solving a weighted least-squares problem \cite{abur2004power}: $\min_V \norm{[z_\text{meas}^T,S_\text{psd}^T]^T - h(V)}_W^2$, where $h(\cdot)$ is a measurement function and $\norm{x}_W^2 = x^T W x$ for a weight matrix $W\succ0$. As proposed in \cite{picallo2017twostepSE}, SE can be decomposed in two parts: First, using the pseudo-measurement $S_\text{psd}$, we solve the power flow to obtain a prior estimate $V_\text{prior}$: 
\begin{equation}\label{eq:PF}\begin{array}{l}
V_\text{prior} = \text{PowerFlow}(S_\text{psd})
\end{array}
\end{equation}
Then, using the real-time PMU measurements $z_\text{meas}$, a posterior solution $V_\text{post}$ can be derived using a linear filter:
\begin{equation}\label{eq:update}\begin{array}{l}
V_\text{post} = V_\text{prior} + K(z_\text{meas}-C_\text{meas}V_\text{prior}) 
\end{array}
\end{equation}
where the gain matrix $K$ is obtained by minimizing the error covariance $\text{tr}(\Sigma_\text{post})= \text{tr}(\mathbb{E}[(V_\text{post}-V)^H(V_\text{post}-V)])$, with $(\cdot)^H$ denoting conjugate transpose:
%\begin{equation}\label{eq:sigmapost}
%\arraycolsep=1pt
%\begin{array}{rl}
%\Sigma_\text{post} =&  \Sigma_\text{prior} + K(\Sigma_\text{meas}+C_\text{meas}\Sigma_\text{prior}C_\text{meas}^H)K^H \\[0.1cm]
%&-KC_\text{meas}\Sigma_\text{prior}-\Sigma_\text{prior}C_\text{meas}^HK^H 
%\end{array}
%\end{equation}
%and thus
\begin{equation}\label{eq:K}
K =  \Sigma_\text{prior}C_\text{meas}^H(C_\text{meas}\Sigma_\text{prior}C_\text{meas}^H+\Sigma_\text{meas})^{-1}
\end{equation}
where $\Sigma_\text{prior}$ and $\Sigma_\text{meas}$ are the expected error covariance of the prior estimate $V_\text{prior}$ and the measurements $z_\text{meas}$ respectively. 

\begin{rem}
Extra measurements from non-synchronized real-time sensors, like  magnitude measurements from SCADA, can be included in the posterior update \eqref{eq:update} for a greater improvement of the posterior estimate $V_\text{post}$ by using the first-order approximation of the measurement function \cite{picallo2017twostepSE}.
\end{rem}

\begin{rem}
As shown in \cite{zhou2006alternative}, splitting the problem in two steps yields the same first-order approximation as solving the problem in one step. Moreover, the posterior minimum-variance estimator using the linear update \eqref{eq:update}, is equal to the maximum-likelihood using a weighted least-squares approach \cite{picallo2017twostepSE}. Therefore, we can conclude that for an SE method that assumes Gaussian noises and performs a maximum likelihood estimation, the posterior covariance will be the same $\Sigma_\text{post}$, and thus the method developed here for optimal sensor placement can be also extended for other SE techniques satisfying these conditions.
\end{rem}

Since $V_\text{post}$ in \eqref{eq:update} is an unbiased estimator, SE accuracy can be defined as a function of the posterior covariance matrix $\Sigma_\text{post}$, which needs to be minimized to improve the SE accuracy. After some manipulations, the error covariance $\Sigma_\text{post}$ for the posterior estimation $V_\text{post}$ can be expressed as:
\begin{equation}\label{eq:sigmaPost2}
\Sigma_\text{post}=(\Sigma_{\text{prior}}^{-1}+C_\text{meas}^H\Sigma_\text{meas}^{-1}C_\text{meas})^{-1} \\[0.1cm]
\end{equation}

Since measurement errors are caused separately by each sensor, they are independent, i.e. $\Sigma_\text{meas}$ is diagonal, and we can split $\Sigma_{\text{post}}$ by every measure:
\begin{equation}\label{eq:sigmaPostfx}\arraycolsep=1pt\begin{array}{l}
\Sigma_{\text{post}}
=(\Sigma_{\text{prior}}^{-1}+\sum_i(C_\text{meas})_{i,\bullet}^H(C_\text{meas})_{i,\bullet}(\Sigma_\text{meas}^{-1})_{i,i})^{-1}\\[0.1cm]
=(\Sigma_{\text{prior}}^{-1}+\sum_ix_i(\tilde{C}_\text{meas})_{i,\bullet}^H(\tilde{C}_\text{meas})_{i,\bullet}(\Sigma_\text{meas}^{-1})_{i,i})^{-1}\\[0.1cm]
\end{array}
\end{equation}
so that $\Sigma_{\text{post}}$ is a function of $x$, where $x_i\in \{0,1\}$, $x_i=1$ if \b{the physical quantity $i$ has a sensor measuring its value}, $0$ otherwise; and $\tilde{C}_\text{meas}$ is the special case of  $C_\text{meas}$ with all possible measurements of all types (bus voltage, bus current, and line current) for all nodes and lines in each phase.%: $\tilde{C}_\text{meas} = [I_\text{d}\;Y\;Y^TA^T]^T$.

\section{Optimal Sensor Placement Problems}\label{sec:optplac}
\vspace{-0.01cm}
In order to improve the accuracy of the SE, the problem of optimal sensor placement consists in minimizing $\Sigma_{\text{post}}(x)$ in \eqref{eq:sigmaPostfx}, according to a metric $m(\cdot)$ and under a set of constraints $h(\cdot)$ to limit the number of sensors or the total cost:
\begin{defi}
Optimal Sensor Placement problem:
\begin{equation}\label{eq:optprob}
\min_x m(\Sigma_{\text{post}}(x)) \text{ s.t. } h(x)\leq 0, \; x_i \in \{0,1\} \; \forall i
\end{equation}
For simplicity, we define: $ f(x)\equiv m(\Sigma_{\text{post}}(x))$.
\end{defi}

\subsection{Metrics for Sensor Placement}\label{sec:metrics}
\vspace{-0.01cm}
There are many possible metrics $f(x)$ available in the context of optimal design of experiments \cite{pukelsheim2006optimal}: 
\begin{itemize}[leftmargin=*]
\item A-\textit{optimal}: $f_\text{A}(x)=\text{tr}(\Sigma_{\text{post}}(x))$. This corresponds to minimizing the sum of the eigenvalues of $\Sigma_{\text{post}}$ and thus the sum of the lengths of the axes of the confidence ellipsoid \cite{boyd2004convex}. This is the metric typically used for SE methods, since standard SE maximizes the log-likelihood through a weighted least-squares minimization \cite{abur2004power}, which is equivalent to the minimum variance estimator using the trace \cite{picallo2017twostepSE}.
\item D-\textit{optimal}: $f_\text{D}(x)=\log\text{det}(\Sigma_{\text{post}}(x))$, where $\text{det}(\cdot)$ is the determinant. This corresponds to minimizing the logarithm of the product of the eigenvalues of $\Sigma_{\text{post}}$ and is related to the logarithm of the volume of the confidence ellipsoid \cite{boyd2004convex}.
\item E-\textit{optimal}: $f_\text{E}(x)=\lambda_\text{max}(\Sigma_{\text{post}}(x))$, where $\lambda_\text{max}(\cdot)$ is the maximum eigenvalue. This corresponds to minimizing the length of the longest axis of the confidence ellipsoid \cite{boyd2004convex}. 
\item M-\textit{optimal}: $f_\text{M}(x) = \max_i(\Sigma_{\text{post}}(x))_{i,i}$, which corresponds to minimizing the highest diagonal term of $\Sigma_{\text{post}}(x)$. 
%This corresponds to minimizing the length of the maximum projection onto the natural basis of the longest axis of $E_\Sigma$. 
\end{itemize}
When relaxing $x_i$ in \eqref{eq:optprob} to be continuous, i.e. $x_i \in [0,1]$, the metrics are all convex in $x$ \cite[Section~7.5]{boyd2004convex} \cite{kekatos2012optimal}.
 
% remark about mixing measurements
%\begin{rem}
%\y{So far we have considered the SE and the optimal sensor placement problems separately, but to be more consistent, they should be solved together:
%\begin{equation}\label{eq:SEnOpt}
%\min_{x,K} m_j(\Sigma_\textup{post}(x,K)) \;
%\end{equation}
%where $m_j$ with $j \in \{\mathrm{A,D,E,M,T}\}$ represents the metric and $\Sigma_\textup{post}(x,K)$ is the posterior covariance as a function of the sensors $x$ deployed and the gain $K$ used  for SE, see \eqref{eq:update}.
%As proven in \cite{picallo2017twostepSE}, SE based on a minimum weighted least-squares yields the same result as the minimum variance SE using the trace as metric, i.e. the A metric. In this case, since we can derive an analytical solution for $K$ \eqref{eq:K}, the problems can be decoupled: 
%\begin{equation}
%\min_x m_\mathrm{A}(\Sigma_\textup{post}(x,\textup{arg}\min_K m_\mathrm{A}(\Sigma_\textup{post}(x,K))))
%\end{equation}
%For $j\neq A$, \eqref{eq:SEnOpt} cannot be decoupled, is too complex to solve, and does not have the desired properties in Table \ref{table:prop}. Therefore, in this work we consider the approximate approach where we use metric A for SE and a general one for the optimal placement:
%\begin{equation}
%\min_x m_j(\Sigma_\textup{post}(x,\textup{arg}\min_K m_\mathrm{A}(\Sigma_\textup{post}(x,K))))
%\end{equation}
%This allows us to take advantage of the properties of the metrics to derive lower and upper bounds on the performance of the optimal solution of the sensor placement problem.}
%\end{rem}

\subsection{Solutions under a budget constraint}
So far we have looked at the cost function in \eqref{eq:optprob}; now we will focus on the constraints: we will consider a budget constraint limiting the total cost of the deployed sensors, which may have a different cost. This constraint allows to take into account the extra cost of installing a sensor in a remote area, or in locations where specific rights might be required, or different types of sensors, etc. This is a more realistic and general approach than using a cardinality constraint, which is the most typical approach in the literature \cite{li2011phasor, kekatos2012optimal, schenato2014bayesian}, and which corresponds to the particular case where sensors have equal costs. The optimal sensor placement problem under a budget constraint can then be expressed as:
\begin{equation}\label{eq:optpmubudget}
x_\text{opt} = \arg\min_x f(x) \text{ s.t. } \sum_i c_i x_i  \leq b,\; x_i\in\{0,1\} \; \forall i
\end{equation}
where $b$ represents the budget and $c_i$ the cost of installing a sensor at location $i$. As mentioned before, computing this optimum is unpractical, and instead suboptimal solutions are typically developed. Therefore, convexity-based bounds are helpful to evaluate the performance of such suboptimal solutions. The relaxed convex problem would be:
\begin{equation}\label{eq:optpmuconvexbudget}
x_\text{convex} = \arg\min_x f(x) \; s.t. \; \sum_i c_i x_i \leq b,\; x_i\in  [0,1]\; \forall i
\end{equation}
We denote the value corresponding to \eqref{eq:optpmuconvexbudget} as $f(x_\text{convex})=f_\text{convex}$. However, $x_\text{convex}$ will not necessarily be feasible to \eqref{eq:optpmubudget}. A feasible solution $x_\text{feas}$ (with respective value $f(x_\text{feas})=f_\text{feas}$) can be built using the convex solution $x_\text{convex}$ of \eqref{eq:optpmuconvexbudget}, by selecting the sensors corresponding to the elements of $x_{\text{convex}}$ with the highest values. This can be done iteratively until the budget is filled, i.e. $\mathcal{B}=\{i\mid c_i \leq b - \sum c_j x_j^{(K-1)} \}=\emptyset$. At every iteration $K$, we would add a new sensor $i^*$ such that:
\begin{equation}\label{eq:optpmufeasbudget}
i^*={\arg\max}_{i \in \mathcal{B} \cup \{i \mid x_i^{(K-1)}=0\}}
 x_{\text{convex},i}, \; x^{(K)} = x^{(K-1)}+e^{i^*}
%k=\underset{i \in \mathcal{B} \cup \{i \mid x_i^{(K-1)}=0\}}{\arg\min}
\end{equation}
where $e^i$ is the vector of the natural basis with $e_i^i=1$, $e_{j \neq i}^i=0$, and $x^{(K)}$ denotes the value of $x$ at iteration $K$ and $x_i^{(0)}=0 \; \forall i$. Ties are broken arbitrarily if there are elements with the same values $x_{\text{convex},i}=x_{\text{convex},j}$ with $i\neq j$. This also applies for further possible ties throughout the paper.

Another simple way to create a feasible solution would be using a forward greedy sensor selection, like the cost-effective algorithm proposed in \cite{leskovec2007cost}, which in every iteration adds the sensor with the lowest ratio of objective function improvement divided by sensor cost:
\begin{equation}\label{eq:optpmugreedybudget}\begin{array}{c}
i^*={\arg\min}_{i \in \mathcal{B} \cup \{i \mid x_i^{(K-1)}=0\}} \frac{f(x^{(K-1)}+e^i)}{c_i}\\ x^{(K)} = x^{(K-1)}+e^{i^*}
\end{array}
\end{equation}
We denote the value corresponding to the final solution $x_\text{greedy}$ of \eqref{eq:optpmugreedybudget} by $f_\text{greedy}$. Then, since $x_\text{opt}$ is a feasible suboptimal solution of \eqref{eq:optpmuconvexbudget}, and $x_\text{feas}$ and $x_\text{greedy}$ are feasible suboptimal solutions of \eqref{eq:optpmubudget}, the following holds for all metrics:
\begin{equation}\label{eq:bound}
f_\text{convex}\leq f_\text{opt} \leq \min(f_\text{greedy},f_\text{feas}) 
\end{equation} 

\section{Optimization Methods}\label{sec:opt}
Since we are considering large networks as the 8500-node feeder in \cite{testfeeder8500}, which translates into having a large number of optimization variables and constraints, standard optimization methods, such as second-order or dual Lagrangian methods, are not well suited to solve the convex problem \eqref{eq:optpmuconvexbudget}. In this section, we take advantage of the structure of the constraints to propose an approach based on first-order projected subgradient methods to get the optimal solution:
\begin{equation}\label{eq:projsubgraddesc}\arraycolsep=1pt\begin{array}{l}
x^{(k+1)} = \Pi_\mathcal{X}(x^{(k)}-\alpha^{(k)}\nabla f(x^{(k)})) \\
\mathcal{X} = \{ x \mid \sum c_i x_i \leq b,\; x_i \in [0,1] \}
\end{array}
\end{equation}
where $\Pi_\mathcal{X}(\cdot)$ denotes the projection on $\mathcal{X}$. We use $\alpha^{(k)} = \frac{\alpha}{k\norm{\nabla f(x^k)}}_2$ to guarantee convergence of the method \cite{nedic2001incremental}, where $\alpha$ is a design parameter.

\subsection{Subgradient Computation}
When developing first-order methods, the gradient expressions are required. The gradients $\nabla f$ for $f \in \{f_\text{A},f_\text{D}\}$ can be derived analytically using matrix calculus \cite{matrixcalc}:
\begin{equation}\label{eq:grad0}\arraycolsep=1pt\begin{array}{rl}
(\nabla f_\text{A}(x))_i = & -\text{tr}\Big(\Sigma_{\text{post}}^2(x)
(\tilde{C}_\text{meas})_{i,\bullet}^H(\tilde{C}_\text{meas})_{i,\bullet}\Big) (\Sigma_\text{meas}^{-1})_{i,i} \\[0.1cm]

(\nabla f_\text{D}(x))_i = & -\text{tr}\Big(\Sigma_{\text{post}}(x)
(\tilde{C}_\text{meas})_{i,\bullet}^H(\tilde{C}_\text{meas})_{i,\bullet}\Big) (\Sigma_\text{meas}^{-1})_{i,i} 
\end{array}
\end{equation}

For $f \in \{f_\text{M}, f_\text{E}\}$ no direct expression for the gradient is available, but the subgradients can be expressed as follows (see Appendix \ref{sec:subgrad}): 
\begin{equation}\label{eq:subgrad0}\arraycolsep=1pt\begin{array}{rl}
(\partial f_\text{M}(x))_i = & 
-\text{tr}\Big(\Sigma_{\text{post}}(x) F^H e_{\max}(x) e_{\max}(x)^T F \Sigma_{\text{post}}(x) \\[0.1cm] &
(\tilde{C}_\text{meas})_{i,\bullet}^H(\tilde{C}_\text{meas})_{i,\bullet}\Big) (\Sigma_\text{meas}^{-1})_{i,i} \\[0.1cm]
(\partial f_\text{E}(x))_i = & 
-\text{tr}\Big(\Sigma_{\text{post}}(x) F^H u_{\max}(x) u_{\max}(x)^H F \Sigma_{\text{post}}(x) \\[0.1cm] &
(\tilde{C}_\text{meas})_{i,\bullet}^H(\tilde{C}_\text{meas})_{i,\bullet}\Big) (\Sigma_\text{meas}^{-1})_{i,i} 
\end{array}
\end{equation}
where $e_{\max}(x) = e^{i^*}$ with the index $i^*$ corresponding to the maximum diagonal entry: $i^*=\arg\max_i(\Sigma_{\text{post}}(x))_{i,i}$; and $u_{\max}(x)= \arg\max_{\norm{u}_2=1} u^H\Sigma_{\text{post}}(x)u$ is the eigenvector corresponding to $\lambda_{\max}(x)$. For a more efficient implementation in terms of number of operations, we rewrite \eqref{eq:grad0} and \eqref{eq:subgrad0} as
\begin{equation}\label{eq:grad}\arraycolsep=1pt\begin{array}{rl}
(\nabla f_\text{A}(x))_i & = - (\tilde{C}_\text{meas})_{i,\bullet} \Sigma_{\text{post}}^2(x)
(\tilde{C}_\text{meas})_{i,\bullet}^H (\Sigma_\text{meas}^{-1})_{i,i} \\[0.1cm]

(\nabla f_\text{D}(x))_i & = - (\tilde{C}_\text{meas})_{i,\bullet} \Sigma_{\text{post}}(x)
(\tilde{C}_\text{meas})_{i,\bullet}^H (\Sigma_\text{meas}^{-1})_{i,i} \\[0.1cm]

(\partial f_\text{M}(x))_i & = - \abs{(\tilde{C}_\text{meas})_{i,\bullet} \Sigma_{\text{post}}(x)F^He_{\max}(x)}^2 (\Sigma_\text{meas}^{-1})_{i,i} \\[0.1cm]

(\partial f_\text{E}(x))_i & = - \abs{(\tilde{C}_\text{meas})_{i,\bullet} \Sigma_{\text{post}}(x)F^Hu_{\max}(x)}^2 (\Sigma_\text{meas}^{-1})_{i,i} 
\end{array}
\end{equation}

\subsection{Projection Algorithm}
Computing the projection $\Pi_\mathcal{X}(\cdot)$ in \eqref{eq:projsubgraddesc} by solving an optimization problem or using the algorithm proposed  in \cite{kekatos2012optimal} may be unpractical for large networks since they may require a large number of iterations, and thus function evaluations. Therefore, we propose a more computationally efficient projection algorithm, in the sense that it is noniterative with a time complexity $\mathcal{O}(n_x)$, where $n_x$ is the dimension of $x$. First, we consider the change of variables $y_i = x_i c_i$ and function $f_c(y)=f\Big(\frac{y}{c}\Big)$, and solve the equivalent problem:
\begin{equation}
\min_y f_c(y) \; s.t. \; \sum_i y_i \leq b,\; y_i\in  [0,c_i]\; \forall i
\end{equation}
using the projected subgradient descent method:
\begin{equation}\label{eq:projsubgraddescy}\arraycolsep=1pt\begin{array}{l}
y^{(k+1)} = \Pi_\mathcal{Y}(y^{(k)}-\alpha^{(k)}\nabla f_c(y^{(k)})) \\
\mathcal{Y} = \{ y \mid \sum y_i \leq b,\; y_i \in [0,c_i] \}
\end{array}
\end{equation}
Then, an algorithm for a computationally efficient projection $\Pi_\mathcal{Y}(\cdot)$ can be derived by modifying the scaled boxed-simplex projection algorithm proposed in \cite{insense2018}. This algorithm consists in finding a $\delta$ and assigning $y_i = \min(\max(z_i-\delta,0),c_i) \; \forall i$, such that $\Sigma y_i \leq b$, see Appendix \ref{sec:appproj}. Then the algorithm can be summarized in the following steps:

\begin{itemize}[leftmargin=*]
\item[1] Null sensors: Find sensor indices $i$ that will have $y_i=0$ after the projection, by sorting $z$ in ascending order and determining the index $i_{0,\geq b}$ of the largest $z_{i_{0,\geq b}}$ such that $\sum_i \min(\max(z_i-z_{i_{0,\geq b}},0),c_i) \geq b$. 
\item[2] Full sensors: Find sensor indices $i$ that will have $y_i=c_i$ after the projection, by sorting $\tilde{z}=z-c$ in descending order and determine the index $i_{1,\leq b}$ of the smallest $\tilde{z}_{i_{1,\leq b}}$ such that $\sum_i \min(\max(z_i-\tilde{z}_{1,\leq b},0),c_i) \leq b$. 
\item[3] Partial sensors: Find the values $y_i$ of sensor indices $i$ that will have $y_i \in (0,c_i)$ after the projection, by computing $\delta$ such that $b=\sum_{\{i\mid z_i \geq \tilde{z}_{1,\leq b}+c_i\}}c_i+ \sum_{\{i\mid z_{0,\geq b}<z_i<\tilde{z}_{1,\leq b}+c_i\}}(z_i-\delta)$: 
\begin{equation}\label{eq:delta}
\delta = \frac{- b + \sum_{\{i\mid z_i \geq \tilde{z}_{i_{1,\leq b}}+c_i\}}c_i + \sum_{\{i\mid z_{i_{0,\geq b}}<z_i<\tilde{z}_{i_{1,\leq b}}+c_i\}}z_i}{\abs{\{i\mid z_{i_{0,\geq b}}<z_i<\tilde{z}_{i_{1,\leq b}}+c_i\}}}
\end{equation} 
and return: $y_i = \min(\max(z_i-\delta,0),c_i) \; \forall i$.
\end{itemize}

\begin{algorithm}[H]\label{alg:alg2}
\caption{$\Pi_\mathcal{Y}$: Projection onto $\mathcal{Y}$}\label{alg:proj}
\begin{algorithmic}[1]
\REQUIRE $b \in \mathbb{R},z,c\in \mathbb{R}^{n_x},z \notin \mathcal{Y}, z_i \geq 0 \; \forall i$
\STATE $z_{i_{0,\geq b}} \hspace{-0.05cm} \leftarrow \hspace{-0.05cm} 0$
\IF{${\{j \hspace{-0.05cm} \mid \hspace{-0.05cm} \sum_i \min(\max(z_i-z_j,0),c_i)\geq b\}} \neq \emptyset$}
	\STATE $z_{i_{0,\geq b}} \leftarrow \max_{\{j \mid \sum_i \min( \max(z_i-z_j,0),c_i)\geq b\}}z_j$ (step 1)
\ENDIF
\STATE $\tilde{z}=z-c$
\STATE $\tilde{z}_{i_{1,\leq b}} \leftarrow \min_{\{j \mid \sum_i \min(\max(z_i-\tilde{z}_j,0),c_i)\leq b\}}\tilde{z}_j$ (step 2)
\IF{$\sum_i \min(\max(z_i-z_{i_{0,\geq b}},0),c_i)= b$}
	\STATE $\delta \leftarrow z_{i_{0,\geq b}}$ (case 1)
\ELSIF{$\sum_i \min(\max(z_i-\tilde{z}_{i_{1,\leq b}},0),c_i)= b$}
	\STATE  $\delta \leftarrow \tilde{z}_{i_{1,\leq b}}$ (case 2) 
\ELSE 
	\STATE $\delta \leftarrow$ \eqref{eq:delta} (case 3, step 3)
\ENDIF
\STATE $y_i \leftarrow \min(\max(z_i-\delta,0),c_i) \; \forall i$
\RETURN  $y$ 
\end{algorithmic}
\end{algorithm}

Note that Algorithm \ref{alg:proj} requires $z_i \geq 0 \; \forall i$, which will be always satisfied since in \eqref{eq:projsubgraddescy} we have $y_i^{(k)}\geq 0$ and $\big(\nabla_y f_c(y) \hspace{-0.2cm} \mid_{y=y^{(k)}} \hspace{-0.25cm} \big)_i\hspace{-0.1cm} = \hspace{-0.1cm} \frac{1}{c_i} \big(\nabla_x f(x) \hspace{-0.2cm} \mid_{x=\frac{y^{(k)}}{c}} \hspace{-0.25cm} \big)_i \hspace{-0.1cm} \leq \hspace{-0.1cm} 0$ due to the expressions in \eqref{eq:grad}. Moreover, Algorithm \ref{alg:proj} has time complexity $\mathcal{O}(n_x)$, since finding $z_{i_{0,\geq b}}$ and $\tilde{z}_{i_{1,\leq b}}$ requires $\mathcal{O}(n_x)$ operations.

\begin{prop} Algorithm \ref{alg:proj} produces the projection $y=\Pi_\mathcal{Y}(z)$ for a $z\notin \mathcal{Y}$ (proof in Appendix \ref{sec:appproj}).
\end{prop}

\section{Test Case}\label{sec:testcase}
\begin{figure*}[!t]
\centering
\subfloat[A-optimal]{\includegraphics[width=8.5cm]{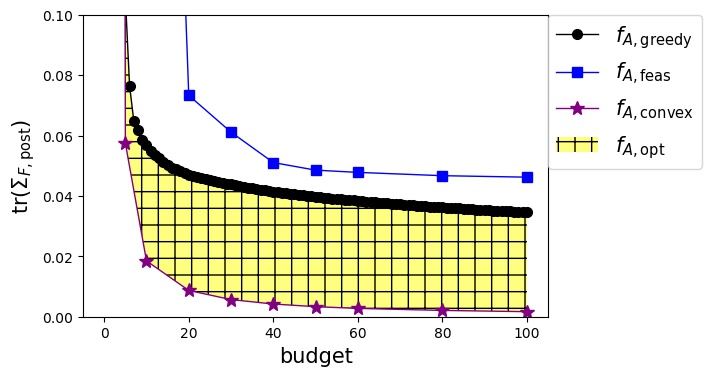}
\label{fig:tracesigmabudget}}
%\hfil
\subfloat[D-optimal]{\includegraphics[width=8.5cm]{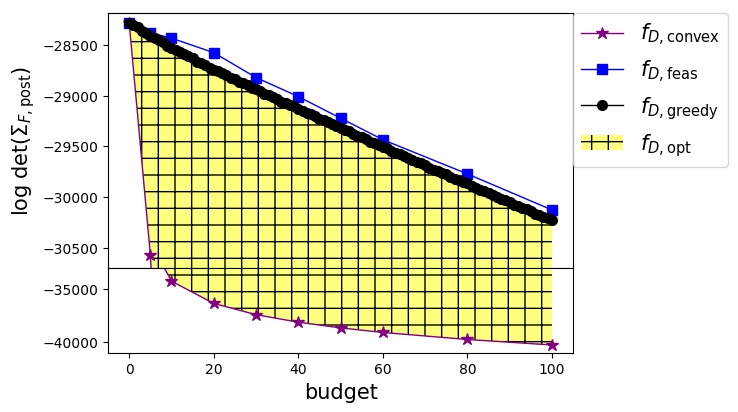}
\label{fig:logdetsigmabudget}}
\hfil
\vspace{-0.15cm}
\subfloat[E-optimal]
{\includegraphics[width=8.5cm]{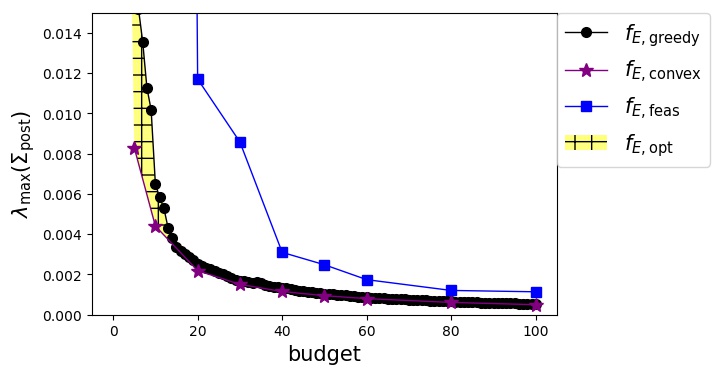}
\label{fig:eigsigmabudget}}
%\hfil
\subfloat[M-optimal]
{\includegraphics[width=8.5cm]{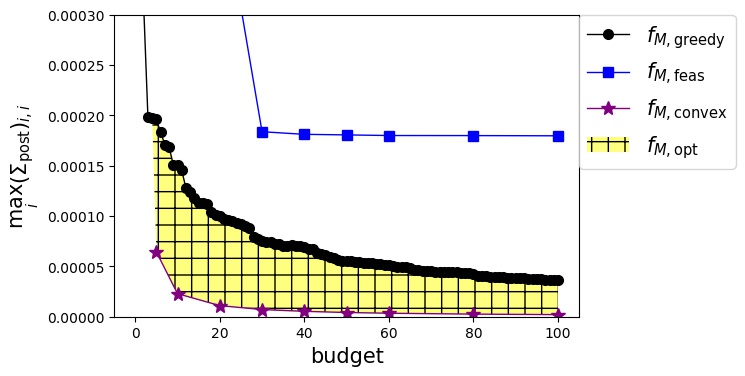}
\label{fig:diagsigmabudget}}
\vspace{-0.1cm}
\caption{Plots for the A,D,E,M-optimal metrics under a budget constraint, showing the lower bounds based on convex relaxations and the upper bounds given by the greedy and feasible solutions. The yellow shaded area with horizontal and vertical lines shows the possible locations of the optimal values $f_{\{\text{A,D,E,M}\},\text{opt}}$.}
\label{fig:results}
%\vspace{-0.5cm}
\end{figure*}

\vspace{-0.01cm}
We have tested the algorithms on the 8500-node test feeder \cite{testfeeder8500} for the different metrics. We assign random normal distributed costs: $c_i \sim \mathcal{N}(1,0.1)$. The algorithms are coded in Python and run on an Intel Core i7-6700HQ CPU at 2.60GHz with 16GB of RAM. Fig. \ref{fig:results} shows the bounds for the 8500-node test feeder. The A,D,E,M-optimal metrics are analyzed under a budget constraint. The yellow shaded area with horizontal and vertical lines shows the area between the minimum upper bound and the maximum lower bound, and thus the possible locations of the optimal values $f_{\{\text{A,D,E,M}\},\text{opt}}$.

It can be observed is that for all metrics, the feasible solutions $f_{\{\text{A,D,E,M}\},\text{feas}}$ \eqref{eq:optpmufeasbudget} perform worse than the greedy ones.
For the A,E,M-optimal metrics the convexity-based bound $f_{\{\text{A,E,M}\},\text{convex}}$ \eqref{eq:optpmuconvexbudget} is relatively close to the greedy solution $f_{\{\text{A,E,M}\},\text{greedy}}$ \eqref{eq:optpmugreedybudget}, especially for the E-optimal metric. This means that the performance of the greedy solutions is close to the performance of the optimal one. 
For the D-optimal metric, the convexity-based bound is far away from the greedy and feasible solutions. Therefore, this bound does not inform about the quality of the greedy and feasible solutions. 
 
\section{Conclusions}\label{sec:conc}
\vspace{-0.01cm}
We have analyzed the problem of optimal PMU placement to minimize the uncertainty of state estimation in distribution grids under a budget constraint. Using the convexity of the metrics considered, we have developed computationally efficient first-order projected subgradient descent methods that combine subgradient expressions with efficient projection algorithms. We have used these optimization methods to solve the convexified optimal sensor placement problem and to derive a lower bound for the value of the optimal solution for large distribution grids. For the A,E,M-optimal metrics, we have observed how suboptimal solutions have a close to optimal performance. 

Future work could include extending these results to take into account network reconfiguration due to switches. Additionally, more exhaustive search algorithms could be developed to obtain solutions with performance closer to the convexity-based lower bound, and in parallel, apply branch and bound methods to push up the convexity-based lower bounds. Moreover, for the D-optimal metric, others bounds based on other properties like supermodularity could be used to obtain tighter bounds \cite{li2011phasor}.
%\fi

\appendices
\section{Proof of Budget Projection}\label{sec:appproj}

\begin{proof}
Note that the projection $y=\Pi_\mathcal{Y}(z)$ is equivalent to solving the optimization problem:
\begin{equation}
\min_{y} \frac{1}{2} \norm{y-z}_2^2 \mbox{ s.t. } \sum y_i \leq b, \; y_i \in [0,c_i]
\end{equation}
with Lagrangian $L = \frac{1}{2} \norm{y-z}_2^2 + \bar{\mu}^T(y-c) + \underbar{$\mu$}^T(-y) + \lambda(\sum y_i-b)$. Here we prove that Algorithm \ref{alg:proj} produces a $y$ that satisfies the Karush-Kuhn-Tucker conditions for some $\bar{\mu},\underbar{$\mu$},\lambda$:
%To prove that a solution $y$ is optimal, it is sufficient to find dual variables $\bar{\mu},\underbar{$\mu$},\lambda$ satisfying the Karush-Kuhn-Tucker conditions:
\begin{equation}\label{eq:KKT}\arraycolsep=1pt\begin{array}{l}
\mbox{stability: } \nabla_{y_i} L = y_i-z_i+\bar{\mu}_i-\underbar{$\mu$}_i+\lambda = 0 \; \forall i\\
\mbox{primal feasibility: } y_i-c_i \leq 0, \; -y_i \leq 0 \; \forall i, \; \sum y_i-b \leq 0 \\
\mbox{dual feasibility: } \bar{\mu}_i \geq 0, \; \underbar{$\mu$}_i \geq 0 \; \forall i, \; \lambda \geq 0\\
\begin{array}{rl}
\mbox{complementary slackness: } &
(y_i-c_i)\bar{\mu}_i = y_i\underbar{$\mu$}_i = 0 \; \forall i \\ & \lambda(\sum y_i-b) = 0
\end{array}
\end{array}
\end{equation} 
According to the complementary slackness and stability conditions we need to have: 
\begin{equation}\label{eq:slack}\arraycolsep=1pt\begin{array}{l}
\mbox{If } y_i=c_i, \mbox{then } \underbar{$\mu$}_i=0, \; \bar{\mu}_i = z_i-c_i-\lambda \\
\mbox{If } y_i=0, \mbox{then } \underbar{$\mu$}_i = \lambda-z_i, \; \bar{\mu}_i=0 \\
\mbox{If } y_i \in (0,c_i), \mbox{then } \underbar{$\mu$}_i=0, \; \bar{\mu}_i = 0, \; y_i = z_i-\lambda\\
\mbox{If } \sum y_i < b, \mbox{then } \lambda = 0 
\end{array}
\end{equation}
Given \eqref{eq:slack}, finding a solution corresponds to finding a $\lambda \geq 0$ so that when assigning $y_i = \min(\max(z_i-\lambda,0),c_i) \; \forall i$, the conditions $\Sigma y_i \leq b, \lambda (\Sigma y_i-b)=0$ are met. The remaining conditions in \eqref{eq:KKT} are satisfied automatically by construction. If Algorithm \ref{alg:proj} falls into cases 1 or 2, the choices for $\lambda$ are simple: $z_{i_{0,\geq b}}$ or $\tilde{z}_{i_{1,\leq b}}$ respectively. Let us consider case 3, which means that there is at least one $y_i \in (0,c_i)$, $y_i <z_i$, so that $\lambda > 0$ and $\sum y_i = b$. First, define the indices of $z$ and $\tilde{z}$ following immediately after $z_{i_{0,\geq b}}$ and $\tilde{z}_{i_{1,\leq b}}$ respectively in Algorithm \ref{alg:proj}: 
\begin{equation}\arraycolsep=1pt\begin{array}{rl}
i_{0,< b} \leftarrow & \arg\min_{\{j \mid z_j > z_{i_{0,\geq b}}\}}z_j \\ 
& = \arg\min_{\{j \mid \sum_i \min(\max(z_i-z_j,0),c_i)< b\}}z_j\\
i_{1,> b} \leftarrow & \arg\max_{\{j \mid \tilde{z}_j < \tilde{z}_{i_{1,\leq b}}\}}\tilde{z}_j \\
& = \arg\max_{\{j \mid \sum_i \min(\max(z_i-\tilde{z}_j,0),c_i)> b\}}\tilde{z}_j
\end{array}
\end{equation}
If falling into case 3, by construction we know that a $\lambda$ satisfying the conditions in \eqref{eq:KKT} will be so that 
\begin{equation}\label{eq:condlambda}
z_{i_{0,\geq b}} < \lambda < z_{i_{0,< b}},\tilde{z}_{i_{1,\leq b}} > \lambda > \tilde{z}_{i_{1,> b}}
\end{equation}
Otherwise we would end in contradictions:
\begin{equation*}\arraycolsep=1pt\begin{array}{rl}
\lambda \leq \tilde{z}_{i_{1,> b}} \implies & b = \sum \min(\max(z_i-\lambda,0),c_i) \\ & \geq \sum \min(\max(z_i- \tilde{z}_{i_{1,> b}},0),c_i) >^{c3} b \\
\lambda \geq \tilde{z}_{i_{1,\leq b}} \implies & b = \sum \min(\max(z_i-\lambda,0),c_i) \\ & \leq \sum \min(\max(z_i- \tilde{z}_{i_{1,\leq b}},0),c_i) <^{c3} b \\
\lambda \geq z_{i_{0,< b}} \implies & b = \sum \min(\max(z_i-\lambda,0),c_i) \\ & \leq \sum \min(\max(z_i- z_{i_{0,< b}},0),c_i) <^{c3} b \\
\lambda \leq z_{i_{0,\geq b}} \implies & b = \sum \min(\max(z_i-\lambda,0),c_i) \\ & \geq \sum \min(\max(z_i- z_{i_{0,\geq b}},0),c_i) >^{c3} b 
\end{array}
\end{equation*}
where the last strict inequalities $<^{c3}, >^{c3}$ are due to being in case 3. Hence, we have
\begin{equation*}\arraycolsep=1pt\begin{array}{rl}
b & = \sum y_i = \sum \min(\max(z_i-\lambda,0),c_i) \\
& \hspace{-0.07cm}\stackrel{\eqref{eq:condlambda}}{=} \sum_{\{i\mid z_i \geq \tilde{z}_{i_{1,\leq b}}+c_i\}}c_i+ \sum_{\{i\mid z_{i_{0,\geq b}}<z_i<\tilde{z}_{i_{1,\leq b}}+c_i\}}(z_i-\lambda) 
\end{array}
\end{equation*}
Therefore we choose $\lambda$ = $\delta$ from \eqref{eq:delta} to satisfy \eqref{eq:KKT}.
\end{proof}

\section{Subgradients of $f_{\{\text{E,M}\}}$}\label{sec:subgrad}
Let us consider the function: $f_u(x) = u^H\Sigma_\text{post}(x)u$, which is convex in $x$. Then the metrics can be reformulated as $f_\text{E}(x)=\max_{\norm{u}_2=1} f_u(x),f_\text{M}(x)=\max_{u \in \{e^1,...,e^N\}}f_u(x)$. Let $x_\text{E}^*, x_\text{M}^*$ denote the optimum $x$ for each metric: $x^*=\arg\min_{c^Tx \leq b} f(x)$. Then we have $\forall \tilde{x}$:
\begin{equation*}\arraycolsep=1pt\begin{array}{rl}
\hspace{-0.05cm} (\nabla_x f_{u_{\max}(\tilde{x})}(x) \hspace{-0.1cm} \mid_{x = \tilde{x}})^T \hspace{-0.05cm} (x_\text{E}^* - \tilde{x}) \hspace{-0.05cm} & 
\stackrel{a)}{\leq} \hspace{-0.05cm} f_{u_{\max}(\tilde{x})}(x_\text{E}^*) \hspace{-0.05cm} - \hspace{-0.05cm} f_{u_{\max}(\tilde{x})}(\tilde{x}) \\ &
\stackrel{b)}{\leq} \hspace{-0.05cm} f_{u_{\max}(x_\text{E}^*)}(x_\text{E}^*) \hspace{-0.05cm} - \hspace{-0.05cm} f_{u_{\max}(\tilde{x})}(\tilde{x}) \\
\hspace{-0.05cm} (\nabla_x f_{e_{\max}(\tilde{x})}(x) \hspace{-0.1cm}\mid_{x = \tilde{x}} )^T \hspace{-0.05cm} (x_\text{M}^* - \tilde{x}) \hspace{-0.05cm} &
\stackrel{a)}{\leq} \hspace{-0.05cm} f_{e_{\max}(\tilde{x})}(x_\text{M}^*) \hspace{-0.05cm} - \hspace{-0.05cm} f_{e_{\max}(\tilde{x})}(\tilde{x}) \\ &
\stackrel{b)}{\leq} \hspace{-0.05cm} f_{e_{\max}(x_\text{M}^*)}(x_\text{M}^*) \hspace{-0.05cm} - \hspace{-0.05cm} f_{e_{\max}(\tilde{x})}(\tilde{x})  
\end{array}
\end{equation*}
where $a)$ holds because $f_u(x)$ is convex in $x \; \forall u$; $b)$ because $u_{\max}(x_\text{E}^*)$ and $e_{\max}(x_\text{M}^*)$ maximize $f_u(x_\text{E}^*)$ and $f_{e^i}(x_\text{M}^*)$ over $u$ and $e^i$ respectively. So $\nabla_x f_{u_{\max}(\tilde{x})}(\tilde{x})$ and $ \nabla_x f_{e_{\max}(\tilde{x})}(\tilde{x})$ are subgradients at $\tilde{x}$ of $f_\text{E}$ and $f_\text{M}$ respectively, and can be computed using matrix algebra \cite{matrixcalc} as in \eqref{eq:grad0}:
\begin{equation*}\arraycolsep=1pt\begin{array}{rl}
(\nabla_x f_u(x) \mid_{x = \tilde{x}})_i = & 
-\text{tr}\Big(\Sigma_{\text{post}}(x) F^H uu^H F \Sigma_{\text{post}}(x) \\[0.1cm] &
(\tilde{C}_\text{meas})_{i,\bullet}^H(\tilde{C}_\text{meas})_{i,\bullet}\Big) (\Sigma_\text{meas}^{-1})_{i,i}
\end{array}
\end{equation*}

\bibliographystyle{IEEEtran}
\bibliography{ifacconf}

\end{document}